\documentclass[a4paper,twocolumn,11pt]{quantumarticle}
\pdfoutput=1
\usepackage[utf8]{inputenc}
\usepackage[english]{babel}
\usepackage[T1]{fontenc}
\usepackage{amsmath}
\usepackage{hyperref}
\usepackage{biblatex}
\usepackage{tikz}
\usepackage{lipsum}
\usepackage{braket}
\usepackage{mathtools}
\usepackage{hyperref}

\begin{document}

\title{Dimensionality reduction with variational encoders based on subsystem purification}

\author[1]{Raja Selvarajan}
\author[2]{Manas Sajjan}
\author[3]{Travis S. Humble}
\author[1,2,4,*]{Sabre Kais}

\affil[1]{Purdue University, Department of Physics and Astronomy, West Lafayette, IN 47907, USA}
\affil[2]{Purdue University, Department of Chemistry, West Lafayette, IN 47907, USA}
\affil[3]{Oak Ridge National Laboratory (ORNL), Oak Ridge, TN 37830}
\affil[4]{Purdue Quantum Science and Engineering Institute, West Lafayette, IN 47907, USA}

\affil[*]{Corresponding author: Sabre Kais, kais@purdue.edu}
\email[*]{kais@purdue.edu}


\maketitle

\begin{abstract}
Efficient methods for encoding and compression are likely to pave way towards the problem of efficient trainability on higher dimensional Hilbert spaces overcoming issues of barren plateaus. Here we propose an alternative approach to variational autoencoders to reduce the dimensionality of states represented in higher dimensional Hilbert spaces. To this end we build a variational based autoencoder circuit that takes as input a dataset and optimizes the parameters of Parameterized Quantum Circuit (PQC) ansatz to produce an output state that can be represented as tensor product of 2 subsystems by minimizing $Tr(\rho^2)$. The output of this circuit is passed through a series of controlled swap gates and measurements to output a state with half the number of qubits while retaining the features of the starting state, in the same spirit as any dimension reduction technique used in classical algorithms. The output obtained is used for supervised learning to  guarantee the  working of the encoding procedure thus  developed. We make use of Bars and Stripes dataset (BAS) for an 8x8 grid to create efficient encoding states and report a classification accuracy of 95\% on the same. Thus the demonstrated example shows a proof for the working of the method in reducing states represented in large Hilbert spaces while maintaining the features required for any further machine learning algorithm that follow. 
\end{abstract}

\section{Introduction}

Variational quantum algorithms  in the NISQ \cite{preskill2018quantum} era provides a promising route towards developing useful algorithms that allow for optimizing states in higher dimensional spaces by tuning polynomial number of parameters. The most prominent techniques within variational methods include Variational Quantum Eigensolver (VQE) \cite{peruzzo2014variational}, Quantum Approximate Optimization Algorithm (QAOA) \cite{farhi2014quantum} and other classical machine learning inspired ones. We ask the readers to refer \cite{D2CS00203E} for an exhaustive study on quantum machine learning with applications in chemistry \cite{sajjan2021quantum, xia2018quantum, kale2021constructive}, physics \cite{selvarajan2022variational},supervised image classification \cite{dixit2021training} and optimization \cite{selvarajan2021prime}. Within the context of optimization and machine learning in general, some of the major problems that needs to be addressed includes  encoding classical data, finding an expressible enough ansatz (Expressibility) \cite{sim2019expressibility}, efficiently computing gradients(Trainability) \cite{du2021learnability}, generalizability \cite{Banchi_2021}. These problems are interlinked and thus not treated independently in general. 

As we move away from the NISQ era towards deep Parameterized Quantum Circuits (PQC), one of the major problems with regards to trainability that needs addressing is the problem of vanishing gradients referred to as barren plateaus \cite{mcclean2018barren}. This might be an affect of working with large number of qubits \cite{mcclean2018barren}, expressive circuit ansatz \cite{holmes2022connecting}, noise induced\cite{wang2007noise} or the use of global cost functions in the learning \cite{cerezo2021cost}. Having efficient procedures to reduce the dimensionality of input quantum state representation will pave a path in developing efficient encoding schemes that could later be used as inputs to other machine learning algorithms where the cost functions on higher dimensional spaces with expressive ansatz are less likely to be trainable. To this end we develop machine learning techniques that allow for compact representations of given input quantum state. 

Within the classical machine learning community autoencoders have been effectively used to develop low dimensional representation of samples generated from a given probability distributions \cite{Kingma_2019}. Inspired from these techniques work on Quantum Autoencoders \cite{romero2017quantum, wan2017quantum} have allowed for people to develop compact representations against a fixed finite state.  It is not clear that such tensor product states with a fixed finite state is always possible and retains the maximal possible information. Here we show that if one were to relax the condition towards maintaining a fixed finite state, a better compact representation can be generated that can be post processed towards classification. We develop techniques to create subsystem purifications for a given set of inputs, and follow it by creating superpositions of these purifications indexed by the subsystem number. This representation is further used for doing classification achieved by applying variational methods over parameterized quantum circuits restricted to this compact representation and show the learning of the method. We apply an ansatz to create subsystem purification on Bars and Stripes (BAS) dataset and show that one can reduce the number of qubits required to represent the data by half and achieve a 95\% classification accuracy on the Bars and Stripes (BAS) dataset. The demonstrated example shows a proof for the working of the method in reducing states represented in large Hilbert spaces while maintaining the features required for any further machine learning algorithm that follow. The scheme thus proposed can be extended to problems with states in large Hilbert spaces where dimensionality reduction plays a key role with regards to the trainability of the parameterized quantum circuit.

\section{Method}

Given an ensemble of input states,  $E =\{\ket{\psi_i}\}$, the objective is to construct a low dimensional representation of states sampled from this distribution $E$. Let $\ket{\psi_i}$ be a state over $n_A + n_B$ qubits. We design a protocol that allows for us to create an equivalent compact representation of $\ket{\psi}$ with $max(n_A,n_B) + 1$ qubits. To simplify the discussion lets assume that $n_A=n_B$ and thus we create a representation using half the qubits. We do this in 2 stages. 

\bigskip
\textbf{Stage 1:}

\begin{figure}
    \centering
    \includegraphics[width=\linewidth]{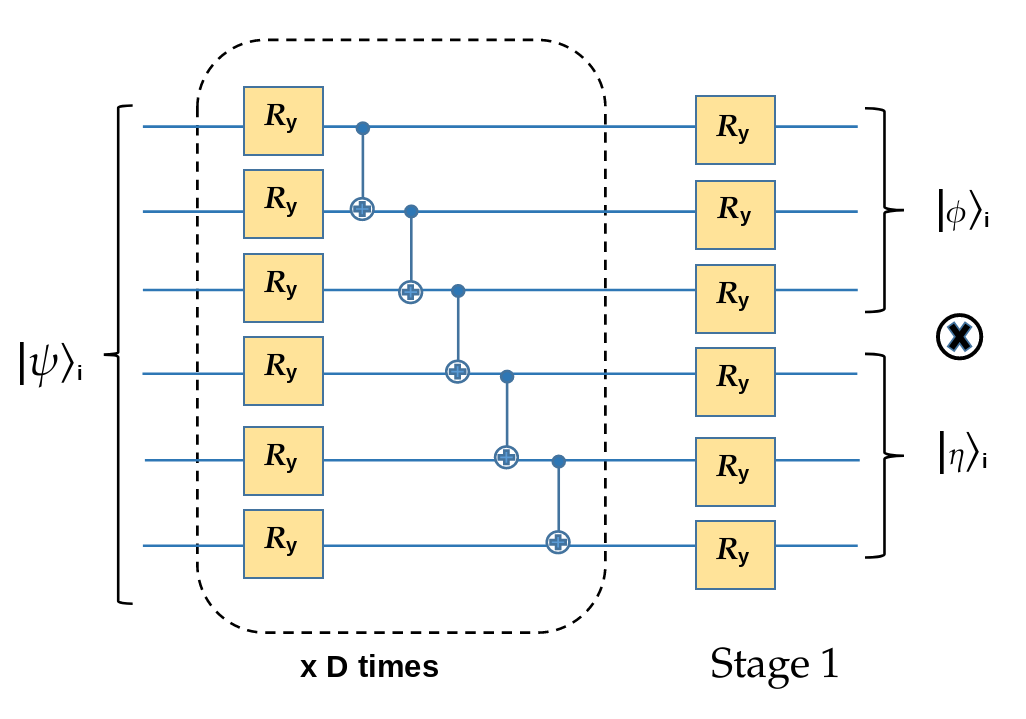}
   \caption{Ansatz used for Encoding circuit $U(\vec{\theta})$ in stage 1. The circuit shows $D$ repeating layers of a unit consisting of $R_y$ gates parameterized by one independent angle each and a ladder of $CNOT$ operators. The circuit is optimized over the dataset to generate equivalent states with subsystem tensor product structure. Thus we obtain $U(\vec{\theta})\ket{\psi}=\ket{\phi}\otimes\ket{\eta}$. $\phi$ is first subsystem and shall be indexed later with an ancilla state $\ket{0}$ and $\ket{\eta}$ is the second subsystem which shall be indexed by $\ket{1}$ }
   \label{fig:variational_ansatz}
\end{figure}

In the first stage we apply a unitary $U(\vec{\theta})$ that decomposes $\ket{\psi_i}_{A,B}$ into $\ket{\alpha(\theta)_i}_{A} \otimes \ket{\beta(\theta)_i}_{B}$. To produce such a tensor product structure we could minimize the entropy on either of subsystems $A$ or $B$ till we get zero. Thus we could optimize over the cost function,

\begin{equation}
   C^{1}_B(\vec{\theta}) = \Bigg< S\bigg(tr_A \Big[U(\vec{\theta})\ket{\psi}_{AB}\bra{\psi}_{AB}U^{\dag}(\vec{\theta})\Big]\bigg) \Bigg>_{\{\ket{\psi}\}}
\end{equation}

where $tr_A$ represents the tracing operation over the qubits of subsystem $A$, $<.>_{\{\ket{\psi}\}}$ represents the averaging over the $\{\ket{\psi}\}$ and $S(\rho) = tr(\rho log(\rho)) $ is the entropy of a given density matrix $\rho$.The cost function $C_B(\vec{\theta})$ attains a maximum value equal to $log(n_B)$ when $\rho_B$ is maximally mixed, and equal to $0$ when $\rho_B$ is a pure state. Fig \ref{fig:variational_ansatz} shows a schematic representation of the ansatz used for $U(\theta)$.

 Variational quantum algorithms have been studied in the past towards creating thermal systems by minimizing the output state against the free energy \cite{Wang_2021,quantarxiv}. The main problem tackled in these papers involves developing techniques that allows one to compute the gradients of Entropy required to be optimized over the training. The issue arises from not having exact representations that can compute logarithm of given density matrix efficiently. Further more to avoid numerical instabilities in the entropy function arising from the density matrix of pure states being singular, here we alternatively maximize over the cost function,

\begin{equation}
\label{eqn:purification}
    C_{AB}(\vec{\theta})) = \Bigg< Tr_A({\rho_A}^2) + Tr_B({\rho_B}^2) \Bigg>_{\{\ket{\psi}\}}
\end{equation}

where $\rho_A = Tr_B(U(\vec{\theta})\ket{\psi}_{AB}\bra{\psi}_{AB}U^{\dag}(\vec{\theta}))$ and $\rho_B = Tr_A(U(\vec{\theta})\ket{\psi}_{AB}\bra{\psi}_{AB}U^{\dag}(\vec{\theta}))$. $C_{AB}$ attains a maximum value 2 when $\rho_A$ or $\rho_B$ are pure states resulting in $Tr(\rho_{A/B}^2) = Tr(\rho_{A/B}) = 1$ and attains a least value $2/n$. 

\begin{figure}
    \centering
    \includegraphics[width=\linewidth ]{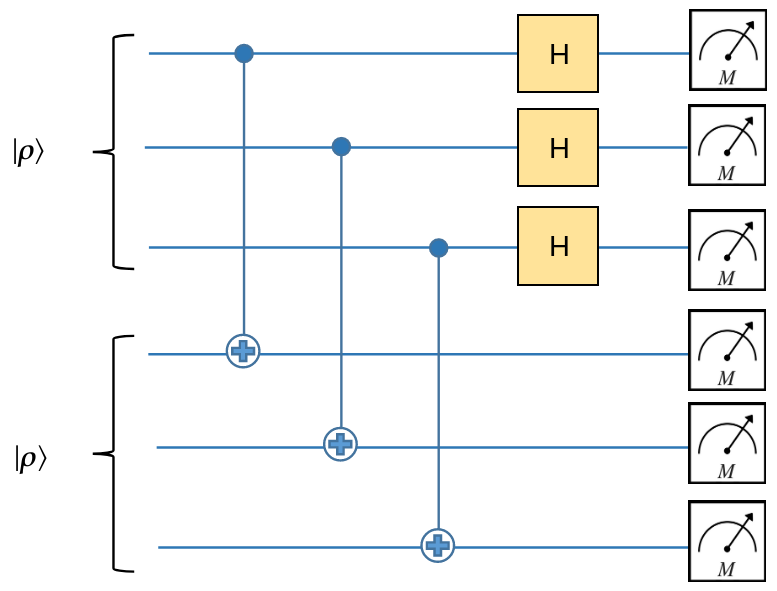}
   \caption{Quantum circuit above implements destructive swap test. Given 2 different density matrices as inputs the circuit computes fidelity of states $F(\gamma,\sigma) = (Tr(\sqrt{\sqrt{\gamma}\sigma\sqrt{\gamma}})^2$, where $\gamma$ and  $\sigma$ are 2 density matrices. Here we use $\gamma = \sigma = \ket{\rho}\bra{\rho}$. Post-processing of measurements with input as 2 copies of $\ket{\rho}$ is used to compute $Tr(\rho^2)$ \cite{cincio2018learning}}
   \label{fig:ansatz}
\end{figure}

The parameters $\vec{\theta}$ are variationally optimized to obtain $\vec{\theta}^{*}=\mathrm{argmax}_{\vec{\theta}} \;\ C_{AB}(\vec{\theta}) $. If $C_{AB}(\vec{\theta})$ reaches an optimal value of zero, we can express $\ket{\psi}_{AB} = \ket{\phi}_{A}\bigotimes\ket{\eta}_{B}$, thus expressing a state with $2^{n+m}$ degrees of freedom effectively using $2^n +2^m$ degrees of freedom. Having expressed the input state as a tensor product of subsystems we now move to stage 2 of the algorithm.

\bigskip
\newpage

\textbf{Stage 2:}

\begin{figure}
    \centering
    \includegraphics[width=\linewidth ]{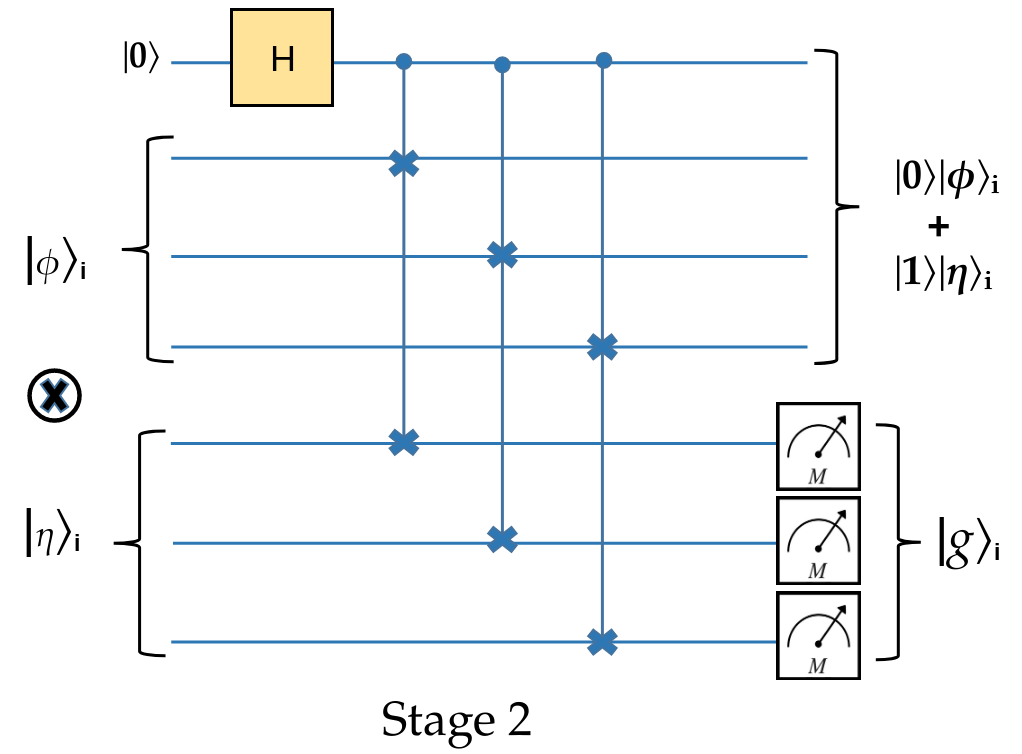}
   \caption{A schematic representation of the steps involved in Stage 2 to prepare the superposition state using an extra ancilla from the product state output of Stage 1. Controlled swap gates are used to generate $\ket{0}\ket{\phi} + \ket{1}\ket{\eta}$. Following this the second subsystem is measured in the computational basis imparting relative phase and amplitude (not shown in the above representation)}
   \label{fig:Stage2}
\end{figure}

Note that the above representation still makes use of $2n$ qubits to capture the features of $\ket{\psi}$. We now show how this representation can be compressed to using $n+1$ qubits. We show how using an additional ancillary qubit, amplitude amplification and projective measurements one can create the state $\ket{0}\ket{\phi} + \ket{1}\ket{\eta}$ starting from $\frac{1}{\sqrt{2}}(\ket{0}+\ket{1})\ket{\phi}\ket{\eta}$. To do this we apply a CSWAP (controlled swap/Fredkin) gate acting on the qubits of system A and B. Thus we get $\ket{0}\ket{\phi}\ket{\eta} + \ket{1}\ket{\eta}\ket{\phi}$. If $\ket{\eta}$ and $\ket{\phi}$ are not orthogonal states, then there exists atleast one basis element $\ket{g}$ in the computational basis with a nonzero coefficient in both these states. Without a loss of generality lets assume that the measurement collapses onto $\ket{g}$ giving raise to $\frac{1}{\sqrt{1+c^2}}(\ket{0}\ket{\phi} + ce^{i\alpha}\ket{1}\ket{\eta})\otimes\ket{g}$, where $c$ and $\alpha$ are real numbers. The factor $ce^{i\alpha}$ is generated from the relative difference in the coefficients of the state corresponding to $\ket{g}$. To ensure that the state collapse to a specific garbage state $\ket{g}$, we could choose $\ket{g}$ to have the maximum probability among all the basis projections. The factor $ce^{i\alpha}$ can now be absorbed as a global normalization if the ancilla register was prepared in the state  $\frac{1}{\sqrt(1+c^2)}(ce^{i\alpha}\ket{0}+\ket{1})$. This ensures that the output of this stage is $\ket{0}\ket{\phi} + \ket{1}\ket{\eta}$. To extend this description to the case when $\ket{\phi}$ and $\ket{\eta}$ are orthonormal, one just needs to apply a transformation controlled on the ancilla register to break this condition. Fig \ref{fig:Stage2} shows a schematic representation of the main steps involved in creating a superposition with the ancilla register being used as an index to the subsystem outputs of Stage 1.


\bigskip

\textbf{Output:}

Thus we have successfully managed to convert the input state $\ket{\psi}$ to $\ket{0}\ket{\phi} + \ket{1}\ket{\eta}$, as required. Note that this procedure is reversible and hence the representation is unique, thus preserving all information content encoded into input state $\ket{\psi}$. To show its reversible, one just needs to take 2 copies of the output state $\ket{0}\ket{\phi} + \ket{1}\ket{\eta}$, measure the corresponding ancilla to project out $\ket{\phi}\ket{\eta}$, and then apply the inverse of $U(\vec{\theta})$ giving back $\ket{\psi}$. Thus the encoding scheme allows for us to create a representation of input state $\ket{\psi}$ with $2n$ qubits into only $n+1$ qubits. This procedure can be repeated iteratively as long as the output state vectors permit a size reduction quantified by the entropy. If repeated $log(n)$ times a $O(log(n))$ size qubit representation of the $n$ qubit state is achievable. Such compact representations are very much reminiscent of the output representation of states on QRAM, where the features are put in a superposition with the index register working as ancillas.

\bigskip


\section{Results}
To demonstrate the working of the method described above, we pick a toy dataset with images of Bars and Stripes (BAS) and build a compact representation of it. The BAS dataset we consider is a square grid with either some columns being only vertically filled (Bars) or some rows being horizontally filled (Stripes) \cite{benedetti2019generative}. One can easily generate such a supervised dataset and realize that the distribution from which these images are sampled has a low entropy characterization. We randomly sample $1000$ data points from a grid size of $16$x$16$ BAS dataset consisting of $131068$ datapoints represented using amplitude encoding on $8$ qubits.

\begin{figure}[!h]
\centering
    \includegraphics[width=\linewidth]{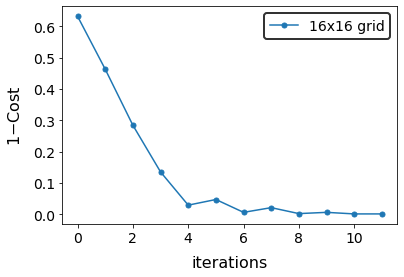}
    \caption{Stage 1: Training cost vs iterations for 16x16 grid.The unitary circuit thus trained creates equivalent tensor product representations using two equal half subsystems of 4 qubits. Note that $1 -$Cost eventually saturates at 0 allowing us to create pure state product subsystem}
    \label{fig:16_encoding_training_a}

  \includegraphics[width=\linewidth]{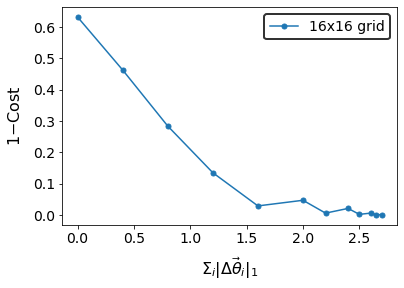}
  \caption{Stage 1: Training cost vs $\sum_i|\Delta\vec{\theta}_i|_1$ for 16x16 grid. The variation in angle as computed by the gradient of eqn \ref{eqn:purification} is minimized as one gets near to the saturation point ($|\Delta{\vec{\theta}|_1}$ measures the 1 norm increase in the angle contribution from the computed gradients with increasing epochs)}
  \label{fig:8_encoding_training_b}
\end{figure}

\begin{figure}[!h]
    \centering
  \includegraphics[width=\linewidth]{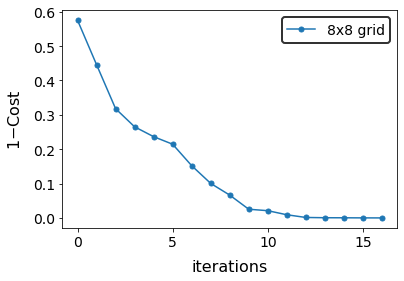}
  \caption{Stage 1: Training cost vs iterations for 8x8 grid. .The unitary circuit thus trained creates equivalent tensor product representations using two equal half subsystems of 3 qubits. Note that $1 -$Cost eventually saturates at 0 allowing us to create pure state product subsystem}
  \label{fig:8_encoding_training_c}

  \includegraphics[width=\linewidth]{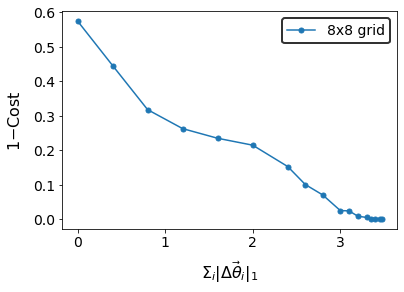}
  \caption{Stage 1: Training cost vs $\sum_i|\Delta\vec{\theta}_i|_1$ for 8x8 grid. The variation in angle as computed by the gradient of eqn \ref{eqn:purification} is minimized as one gets near to the saturation point ($|\Delta{\vec{\theta}|_1}$ measures the 1 norm increase in the angle contribution from the computed gradients with increasing epochs)}
  \label{fig:8_encoding_training_d}
\end{figure}


 Applying the protocol described above we reduce the representation of the state into a tensor product of 2 subsystem of equal sizes. Fig \ref{fig:16_encoding_training_a},\ref{fig:8_encoding_training_c} shows the learning of optimal parameters $\vec{\theta}$ as the cost function falls.  We use standard gradient descent \cite{Wierichs_2022} approach in doing the training . Note the cost function drops to zero implying that the representation thus created is exact with a lossless transformation created by $U(\vec{\theta})$. For the 16x16 grid case, the ansatz $U(\vec{\theta})$ is made of D=5 layers, while for the 8x8 grid is made of D=3 layers. At this point we apply a layer of swap gates to reduce the 8 qubit representation of 16x16 grid samples into 5 qubits and the 6 qubit representation of 8x8 grid samples into 4 qubits.





\begin{figure}[!h]
        \centering
        \includegraphics[width=\linewidth]{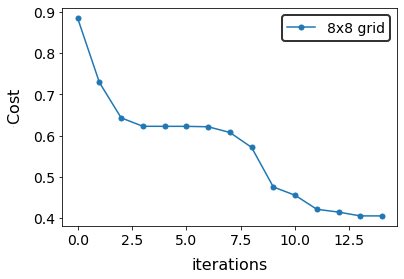}
        \caption{Classification cost vs iterations for 8x8 grid.Figure shows the saturation of classification cost as per eqn \ref{eqn:class_cost} after 13 iterations. }
        \label{fig:classification_training}

       \includegraphics[width=\linewidth]{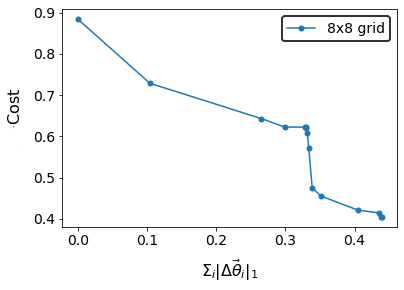} 
       \caption{Classification cost vs $\sum_i|\Delta\vec{\theta_i}|_1$ for 8x8 grid.  Figure shows that the variation in angle as computed by the gradient of eqn \ref{eqn:class_cost} is minimized as one gets near to the saturation point. ($|\Delta{\vec{\theta}|_1}$ measures the 1 norm increase in the angle contribution from the computed gradients with increasing epochs }
       \label{fig:cost_theta}
\end{figure}

We now use this as input for doing supervised classification. We use approx 80\% of the samples from the output of the encoded samples for training and keep the remaining 20\% of the samples for testing. An ansatz $V(\vec{\theta})$ with the same number of qubits as that of the input samples is trained, with the sign of expectation value of pauli-$Z$ operator being used as a label for differentiating between bars and stripes. Input image is classified as a bars image if the expectation value is positive, and stripes image if negative. We use the sum of $2$ norm errors over the dataset labels ($1$ for bars and $-1$ for stripes) as the cost function to be minimized over, i.e,

\begin{equation}
\label{eqn:class_cost}
    \mathrm{cost} = 
\sum_{i} \;\; (l_i - \bra{\Tilde{\psi_i}} V^{\dag}(\vec{\alpha}) [Z\otimes I^{\otimes n-1} ] V(\vec{\alpha})\ket{\Tilde{\psi_i}})^2 
\end{equation}
   
where the summation index $i$ labels the dataset, $l_i$ refers to the labels corresponding to the sample input and $\ket{\psi_i}$ is used to denote the compact representation of the state that the above encoding scheme provides.
For the 8x8 grid, a total 508 bars and stripe images are produced with half of them belonging to each category. We use 400 of these samples for training and 108 samples for testing. Fig \ref{fig:classification_training} shows the cost of optimizing the parameters of $V(\vec{\theta})$ as a function of the number of iterations. We get a $95\%$ accuracy on the testing data, showing that the method use to generate the compact representation did not destroy the features of the input state.

\section{Runtime analysis of Encoding scheme}
Here we shall analytically compute the required runtime for the above described protocol. Lets assume that the input ensemble of $N$ quantum states over $n$ qubits supports a compact representation, allowing us to use the above protocol to encode with half the number of qubits. Let our ansatz to be optimized be made of $d$ layers. Thus stage 1 involves optimization over $2ndN$ parameters for $N$ samples. Using destructive swap test to compute fidelity with an error $\epsilon$ we would require $O(1/\epsilon^2)$ samples. Thus the runtime complexity scales evaluating $O(ndN/\epsilon^2)$ quantum circuits per iteration for Stage 1. Stage 2 involves projection onto the state with largest overlap. The overlap achievable onto any given computational basis $\ket{g}$ can be maximized using grovers with the worst case runtime of $O(2^{n/2})$ steps with query complexity of $O(1/\epsilon)$. Thus the overall runtime is bounded by $O(N(Tnd/\epsilon^2 + 2^{n/2}/\epsilon))$, where $T$ is the number of iterations required in stage 1 optimization. In contrast the runtime of a classical autoencoder to prepare a compact state is $O(NTd2^n)$. We show in the appendix \ref{appendix:A}, how for certain cost functions, using a specialized ansatz and carefully prepared index registers one can get around the exponential cost incurred in preparing the compact superposition state for machine learning tasks.


\section{Discussion and Conclusion}
We discuss a scheme that allows for a compact representation of states in higher dimensional Hilbert spaces using half the number of qubits. The output thus created serves as good starting states for any further machine learning algorithm that might follow. The protocol is based on designing a quantum circuit that allows creating tensor product subsystems and demonstrate results on bars and stripes datasets for 8x8 grid and 16x16 grid. We further use this output to create compact representations with half the number of qubits as compared to the starting state. To show that this representation is a lossless encoding we use it to do supervised learning using variational circuits on the entire dataset of 8x8 grid and reproduce a 95\% accuracy on the training dataset (consisting of 108 samples). Unlike quantum autoencoders where the compact representations rely on being able to optimize against a fixed garbage state, here the relaxed restriction on the tensor product helps provide compact representations in cases where a fixed garbage state would not be feasible. Further investigations on what the entanglement of the subsystems reveal about the probability distribution from which the data is sampled can lead to other useful applications of this protocol. One might also be interested in carrying out machine learning by using weighted quantum circuits that run on the subsystems independently and compare its performance against the compact representations created thereby. One can also imagine using low entropic entangled states that stage 1 protocol outputs as input states for entanglement forging \cite{eddins2022doubling} and look for useful applications with the same.  We would like to conclude by saying that, efficient methods for encoding and compression are likely to pave way towards the problem of efficient trainability on higher dimensional Hilbert spaces, and this work serves as a step towards that direction.

\section{Acknowledgements}

This material is based upon work supported by the U.S. Department of Energy, Office of Science, National Quantum Information Science Research Centers, Quantum Science Center. This manuscript has been authored by UT-Battelle, LLC, under contract DE-AC05-00OR22725 with the US Department of Energy (DOE). Sabre Kais would like to acknowledge the support from National Science Foundation under award number 1955907.

\printbibliography

\onecolumn\newpage
\appendix

\section{Appendix}
\label{appendix:A}

We shall show here that as far as the variational circuit is considered, the presence of phase in the index qubit can be eliminated by the choice of ansatz and the relative probabilities of $\ket{\phi}$ and $\ket{\eta}$ can be ignored with sufficient samples as they average out to be equal.

Let $V(\vec{\alpha})$ be the ansatz used for classification post creating the compact state representation. The classification cost function with respect to this ansatz is given by,

\begin{equation}
\label{eqn:c_cost}
     \mathrm{Classification \; Cost} = \sum_{i} \;\; (l_i - \bra{\Tilde{\psi_i}} V^{\dag}(\vec{\alpha}) [I^{\otimes n-1}\otimes Z  ] V(\vec{\alpha})\ket{\Tilde{\psi_i}})^2 
\end{equation}

where, $\ket{\Tilde{\psi_i}} = c_0\ket{0}\ket{\phi} + c_1 e^{i\gamma}\ket{1}\ket{\eta}$, where $c > 0$ references the relative amplitude (s.t $c_0^2+c_1^2 =1$) and $\gamma$ the relative phase acquired in measuring subsystem 2 after the sequence of controlled swaps performed. We start with creating an ansatz that decouples the effect of having a phase on the state. To this end, we modify the unitary ansatz $V$ to have the following form,  $V(\vec{\alpha_1},\vec{\alpha_2})= \ket{0}\bra{0}\otimes V_0(\vec{\alpha}_0) + \ket{1}\bra{1}\otimes V_1({\vec{\alpha}_1})$, i.e, $V_1(\vec{\theta_1})$ acts only on $\ket{\phi}$ and $V_2(\vec{\theta_2})$ acts only on $\ket{\eta}$. Simplifying \ref{eqn:c_cost} we get,

\begin{equation}
    \mathrm{Classification \; Cost} = \sum_{i} \;\; (l_i  - (c_0^2\bra{\phi}V_0^{\dag}(\vec{\alpha_0})[I^{\otimes n-2}\otimes Z  ]V_0(\vec{\alpha_0})\ket{\phi} + c_1^2\bra{\eta}V_1^{\dag}(\vec{\alpha_1}) [I^{\otimes n-2}\otimes Z  ]V_1(\vec{\alpha_1})\ket{\eta} ))^2  
\end{equation}

 Notice that the effect of having any phase is lost in the process as the ansatz and measurement is impervious to the presence of phase in the state. Now we will show how averaging over multiple realizations, its possible to get rid of the relative probabilities. The gradients computed from the above cost function computed for input sample $i$ is given by,
 
 \begin{equation}
     \nabla \mathrm{Cost}_i = \braket{c_{i,0}^2}\nabla\bra{\phi}V_0^{\dag}(\vec{\alpha_0})[I^{\otimes n-2}\otimes Z  ]V_0(\vec{\alpha_0})\ket{\phi} + \braket{c_{i,1}^2}\nabla\bra{\eta}V_1^{\dag}(\vec{\alpha_1}) [I^{\otimes n-2}\otimes Z  ]V_1(\vec{\alpha_1})\ket{\eta}
 \end{equation}

where, $\braket{.}$ refers to averaging over multiple shots of the same input sample after the measurement of the second subsystem has been made. An estimate made of $\braket{c_{i,0}^2}$ and $\braket{c_{i,1}^2}$ can be estimated by measuring the index register following the measurement of subsystem 2 in the computational basis. We can thus prepare the index register in the state $\frac{\braket{c_{i,1}^2}}{(\braket{c_{i,0}^2}+\braket{c_{i,1}^2})}\ket{0}+\frac{\braket{c_{i,0}^2}}{(\braket{c_{i,0}^2}+\braket{c_{i,1}^2})}\ket{1}$ for measuring the gradient contribution from input sample $i$. This allows us to optimize for the gradients independent of the computational basis onto which system 2 gets projected, giving us,

 \begin{equation}
     \nabla \mathrm{Cost}_i = \sum_i \nabla\bra{\phi}V_0^{\dag}(\vec{\alpha_0})[I^{\otimes n-2}\otimes Z  ]V_0(\vec{\alpha_0})\ket{\phi} + \nabla\bra{\eta}V_1^{\dag}(\vec{\alpha_1}) [I^{\otimes n-2}\otimes Z  ]V_1(\vec{\alpha_1})\ket{\eta}
 \end{equation}

The gradients computed to this cost function is deterministic removing any probabilistic effects from the presence of a relative phase or amplitude within the cost function. This helps us circumvent the exponential scaling in the runtime that might arise from the need to project onto a specific computational basis state.

\end{document}